\documentclass[pre,aps,showkeys,amsmath,amssymb,twocolumn,superscriptaddress]{revtex4}

\usepackage{graphicx}

\begin{document}

\title{Electromagnetic Oscillations in a Spherical Conducting Cavity\\
with Dielectric Layers. Application to Linear Accelerators}

\author{W{\l}adys{\l}aw \.Zakowicz}
\email[]{wladyslaw.zakowicz@ifpan.edu.pl}
\affiliation{Institute of Physics, Polish Academy of Sciences,
Al.~Lotnik\'ow 32/46, 02--668 Warsaw, Poland}

\author{Andrzej A. Skorupski}
\email[]{askor@fuw.edu.pl}
\affiliation{Department of Theoretical Physics,
National Centre for Nuclear Research,
Ho\.za 69, 00--681 Warsaw, Poland}

\author{Eryk Infeld}
\email[]{einfeld@fuw.edu.pl}
\affiliation{Department of Theoretical Physics,
National Centre for Nuclear Research,
Ho\.za 69, 00--681 Warsaw, Poland}

\begin{abstract}
We present an analysis of electromagnetic oscillations in a
spherical conducting cavity filled concentrically with either dielectric or
vacuum layers. The fields are given analytically, and the resonant frequency
is determined numerically. An important special case of a spherical
conducting cavity with a smaller dielectric sphere at its center is treated in
more detail. By numerically integrating the equations of motion we demonstrate
that the transverse electric oscillations in such cavity can be used to accelerate
strongly relativistic electrons. The electron's trajectory is assumed to be nearly tangential
to the dielectric sphere. We demonstrate that the interaction of such electrons with
the oscillating magnetic field deflects their trajectory from a straight
line only slightly. The Q factor of such a resonator only
depends on losses in the dielectric. For existing ultra low loss dielectrics, Q can be
three orders of magnitude better than obtained in existing cylindrical cavities.
\end{abstract}


\maketitle

\section{Introduction}

It has been shown \cite{wz,wza,wz1} that, if a plane electromagnetic wave is
scattered on a finite dielectric object, structural resonances can be excited in the
object (e.g., whispering gallery modes). They are associated with very high
amplitudes of oscillating EM fields in the dielectric and its vicinity. Their maxima
exceed values reached in resonant cavities of typical
linear accelerators by several orders of magnitude. Therefore, one can think of applying these
fields to  accelerate charged particles \cite{wz,wza,wz1}. Many other applications of the
whispering gallery modes are described in \cite{orni,lisey,ilchen}.

As for the proposals given in \cite{wz,wza,wz1}, both light produced by lasers and
microwaves are conceivable.  However, it is difficult to achieve the required
synchronization of wave particle in the optical frequency range. In the microwave frequency range,
this mechanism would require excessive total excitation energy and so may not be practical.

\begin{figure}[t]
\centering\includegraphics[scale = .7]{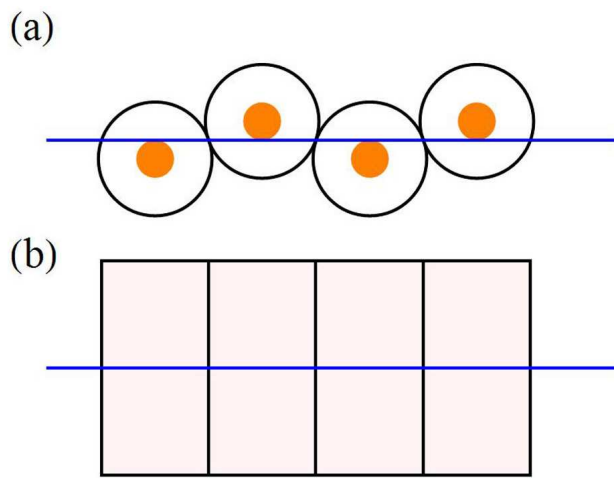} \caption{Proposed
multi cell accelerator units (a) vs. those of SLAC (b), assuming
the same electron transit time through the cavity.}\label{accunits}
\end{figure}

In this paper we demonstrate that the last mentioned problem can be overcome by
locating the dielectric object in a resonant cavity. This appeals to traditional
accelerating structures used in SLAC, see \textbf{Figure~1}. In the latter case,
maximum amplitudes of accelerating fields are restricted by Joule heating losses in
conducting walls and electric breakdown. In this connection, in existing
accelerators (e.g., in LHC) one avoids sharp edges of the walls and uses
superconductive resonant cavities. Unfortunately, since superconductivity of the
walls disappears if the magnetic field on the wall exceeds a critical value, the
maximal values of accelerating fields in these highly complicated cavities are
not much higher than those reached in SLAC.

The presence of a dielectric in the central part of the resonance cavity shifts
the magnetic field maximum from regions close to the metallic wall towards the
dielectric surface. This considerably lowers skin effect losses
in the wall. Even though additional losses due to dielectric heating are
introduced, total losses would nevertheless be lower if one could apply ultra
low loss dielectrics (with $\tan\delta \sim 10^{-7}$) as described in
\cite{taber,krupka}.
In that case, a resonator quality reaching $Q \sim 10^7$ can be obtained, as
compared to $Q \sim 10^4$ for SLAC.

\section{A spherical conductive cavity with dielectric layers}

Our approach to describe electromagnetic oscillations in a resonant cavity
assumes that the cavity can be divided into regions in which the fields can be
determined analytically. The resonant frequency is defined by the fact that the
fields must satisfy boundary conditions at the cavity wall along with continuity
conditions at the interfaces. This frequency will be determined by numerically
solving the consistency condition for these requirements.

\begin{figure}[h]
\centering\includegraphics[scale=.6]{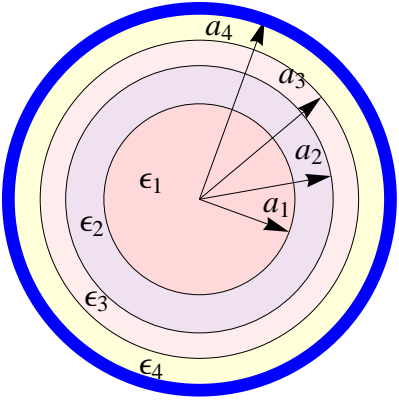}%
\caption{An example of a spherical cavity with dielectric layers ($N=4$).}
\label{cavity}
\end{figure}

In general, we assume that the cavity is bounded by a conducting spherical
surface, and filled concentrically with $N\ (\geq 1)$ either dielectric or vacuum
layers. Each dielectric layer is assumed to be homogeneous. We
introduce a spherical coordinate system $(r, \theta, \phi)$ with its origin at
the cavity center. The layers are bounded by $r = a_1,\ a_2, \ldots ,\ r_{N-1}$,
up to $r = a_N \equiv b$ for the metallic boundary, see \textbf{Figure \ref{cavity}}.

The harmonically oscillating electromagnetic fields in each concentric
layer are described by Maxwell's equations (Gaussian units, magnetic permeability
$\mu=1$, and complex fields proportional to exp$(-i\omega t)$):
\begin{equation}
\nabla \times \mathbf{E} =  ik \mathbf{B}, \quad
\nabla \times \mathbf{B} = -ik \epsilon \mathbf{E}
\label{maxwel}
\end{equation}
where
\begin{equation}
\label{kvom}
k = \omega /c
\end{equation}
$\omega$ is the angular frequency, and $\epsilon$ denotes complex dielectric
permittivity
\begin{equation}
\epsilon = \epsilon' + i \, \epsilon'', \quad | \epsilon'' | \ll \epsilon'
\end{equation}
These fields split into transverse electric (TE) or transverse 
magnetic (TM), which have no radial components of either field \cite{jackson}.
In an ideal resonator with perfectly conducting walls and perfect dielectrics,
pure TE or TM modes can be excited. They will also be approximately valid in real
resonators if their energy losses are not too high.

Using (9.116), and (9.119) in \cite{jackson}, which describe the vacuum TE
field in spherical coordinates, and replacing there $k \to \sqrt{\epsilon} k$ we
obtain the most general form of the TE field in the uniform dielectric:
\begin{equation}\label{Elm}
\begin{split}
\mathbf{E}^{lm} (\mathbf{r}, t) &= \mathbf{E}^{lm} (\mathbf{r},\omega)
e^{-i\omega t}\\
\mathbf{E}^{lm} (\mathbf{r},\omega) &= \mathcal{E}_{\text{t}}(kr)
\mathbf{X}^{lm}(\theta, \phi)
\end{split}
\end{equation}
where 
\begin{equation}
\mathcal{E}_{\mathrm{t}}(r) =
A_l^{(1)} j_l(\sqrt{\epsilon} kr) +
A_l^{(2)} y_l(\sqrt{\epsilon} kr)
\label{Et}
\end{equation}
$\mathbf{X}_{lm}(\theta, \phi)$ are vector spherical harmonics as defined
by Eq. (9.119) in \cite{jackson}, and $j_l(\rho) \equiv
\sqrt{\frac{\pi}{2\rho}}\,J_{l+\frac{1}{2}}(\rho)$
and $y_l(\rho) \equiv\sqrt{\frac{\pi}{2 \rho}}Y_{l+\frac{1}{2}}(\rho)$ are
spherical Bessel and Neumann  functions.

The corresponding magnetic induction can be determined from the first
Maxwell equation (\ref{maxwel}). Using also (10.60) in \cite{jackson} we
obtain
\begin{equation}
\mathbf{B} \equiv \mathbf{B}_{lm} (\mathbf{r}, t) = \mathbf{B}_{lm}
(\mathbf{r},\omega) e^{-i\omega t}
\end{equation}
where $\mathbf{B}_{lm} (\mathbf{r},\omega)$ involves both the transverse
radial component:
\begin{equation}
\mathbf{B}_{lm} (\mathbf{r},\omega) =
\mathbf{B}_{lm\, \text{t}} (\mathbf{r},\omega) +
\mathbf{B}_{lm\, \text{r}} (\mathbf{r},\omega)
\label{Blm}
\end{equation}
in which
\begin{align}
&\mathbf{B}_{lm\, \text{t}} (\mathbf{r},\omega) = 
\mathcal{B}_{\text{t}}(r)\mathbf{n} \times \mathbf{X}_{lm}(\theta, \phi)
\label{Blmt}\\
&\mathbf{B}_{lm\, \text{r}} (\mathbf{r},\omega)
= \mathcal{B}_{\text{r}}(r) Y_{lm}(\theta, \phi) \mathbf{n}
\label{Blmr}\\
&\mathcal{B}_{\text{t}}(r) = - \frac{i}{kr} \Bigl[
A_l^{(1)} j^{\scriptscriptstyle\mathrm{D}}_l(\sqrt{\epsilon} kr) +
A_l^{(2)} y^{\scriptscriptstyle\mathrm{D}}_l(\sqrt{\epsilon} kr) \Bigr]\\
&\mathcal{B}_{\text{r}}(r) = 
\frac{\sqrt{l(l + 1)}}{kr} \mathcal{E}_{\mathrm{t}}(r)\label{Br}
\end{align}
Here $Y_{lm}(\theta, \phi)$ are spherical harmonics, $\mathbf{n} = \mathbf{r}/r$,
$l$ is a positive integer related to the integer $m$ by $-l \leq m \leq l$,
$j^{\scriptscriptstyle\mathrm
{D}}_l(\rho) \equiv \frac{d}{d \rho}(\rho j_l(\rho))$
and $y^{\scriptscriptstyle\text{D}}_l(\rho) \equiv \frac{d}{
d \rho} (\rho y_l(\rho))$ are derivatives of the Riccati--Bessel
and Riccati--Neumann functions.

In a similar way, using (9.118), (9.119) and (10.60) in \cite{jackson}
along with the second Maxwell equation (\ref{maxwel}), we obtain for the
TM modes in the uniform dielectric:
\begin{equation}\label{barBlm}
\begin{split}
\bar{\mathbf{B}}_{lm} (\mathbf{r}, t) &=
\bar{\mathbf{B}}_{lm} (\mathbf{r},\omega)
e^{-i\omega t}\\
\bar{\mathbf{B}}_{lm} (\mathbf{r},\omega) &= \bar{\mathcal{B}}_{\mathrm{t}}(r)
\mathbf{X}_{lm}(\theta, \phi)
\end{split}
\end{equation}
where 
\begin{equation}
\bar{\mathcal{B}}_{\mathrm{t}}(r) =
\bar{A}_l^{(1)} j_l(\sqrt{\epsilon} kr) +
\bar{A}_l^{(2)} y_l(\sqrt{\epsilon} kr)
\label{barBt}
\end{equation}
\begin{equation}
\bar{\mathbf{E}} \equiv \bar{\mathbf{E}}_{lm} (\mathbf{r}, t) =
\bar{\mathbf{E}}_{lm} (\mathbf{r},\omega) e^{-i\omega t}
\end{equation}
where $\bar{\mathbf{E}}_{lm} (\mathbf{r},\omega)$ involves both the transverse, and
radial component:
\begin{equation}
\bar{\mathbf{E}}_{lm} (\mathbf{r},\omega) =
\bar{\mathbf{E}}_{lm\, \text{t}} (\mathbf{r},\omega) +
\bar{\mathbf{E}}_{lm\, \text{r}} (\mathbf{r},\omega)
\label{barElm}
\end{equation}
in which
\begin{align}
&\bar{\mathbf{E}}_{lm\, \text{t}} (\mathbf{r},\omega) = 
\bar{\mathcal{E}}_{\text{t}}(r)
\mathbf{n} \times \mathbf{X}_{lm}(\theta, \phi)\\
&\bar{\mathbf{E}}_{lm\, \text{r}} (\mathbf{r},\omega) =
\bar{\mathcal{E}}_{\text{r}}(r)
Y_{lm}(\theta, \phi) \mathbf{n}\\
&\bar{\mathcal{E}}_{\text{t}}(r) = \frac{i}{kr\epsilon} \Bigl[
\bar{A}_l^{(1)} j^{\scriptscriptstyle\mathrm{D}}_l(\sqrt{\epsilon} kr) +
\bar{A}_l^{(2)} y^{\scriptscriptstyle\mathrm{D}}_l(\sqrt{\epsilon} kr)
\Bigr]\label{barElmt}\\
&\bar{\mathcal{E}}_{\text{r}}(r) =
-\frac{\sqrt{l(l + 1)}}{kr\epsilon} \bar{\mathcal{B}}_{\mathrm{t}}(r)
\label{barElmr}
\end{align}

In our analysis we will admit small energy losses in both the wall and the dielectric
layers. However, when calculating the resonant frequency of the cavity $\omega_0
\equiv \omega^l$, these losses will be neglected. Thus the wall is
assumed to be perfectly conducting. This means that it carries no electric or magnetic
field. Then continuity of the tangential components of the electric field and the
normal components of the magnetic induction at each interface require vanishing of
these components at the boundary of $N$th layer at $r=b$.
In view of (\ref{Elm})--(\ref{Blmr}), and (\ref{barBlm})--(\ref{barElmr})
this will be the case if the following boundary conditions at $r=b$ are fulfilled:
\begin{equation}\label{ANb}
\begin{split}
A_{lN}^{(1)} j_l(\sqrt{\epsilon_{\scriptstyle N}} kb) +
A_{lN}^{(2)} y_l(\sqrt{\epsilon_{\scriptstyle N}} kb)& = 0\\
\bar{A}_{lN}^{(1)} j^{\scriptscriptstyle\mathrm{D}}_l(\sqrt{\epsilon_N} kb) +
\bar{A}_{lN}^{(2)} y^{\scriptscriptstyle\mathrm{D}}_l(\sqrt{\epsilon_N} kb)& = 0
\end{split}
\end{equation}
where the upper line refers to TE modes and the lower one to TM modes.

In the first layer which contains the origin $r=0$ we must choose
\begin{equation}
A_{l1}^{(2)} = 0 \quad \text{or} \quad  \bar{A}_{l1}^{(2)} = 0
\label{A1eq0}
\end{equation}
to avoid singularities of $y_l(\rho)$ or $y^{\scriptscriptstyle\mathrm{D}}_l(\rho)$
at $\rho = 0$.

In the simplest case of a spherical cavity filled completely with the
dielectric (or vacuum), i.e., $N=1$, the boundary conditions (\ref{ANb})
lead to
\begin{equation}
j_l(\sqrt{\epsilon_1} kb) = 0 \quad \text{or} \quad
j^{\scriptscriptstyle\mathrm{D}}_l(\sqrt{\epsilon_1} kb) = 0
\label{jleq0}
\end{equation}
where $\epsilon_1 \geq 1, \  k = \omega_0/c.$
This defines the resonant frequency $\omega_0$ of either TE or TM modes,
depending only on $\epsilon_1$ ($=\epsilon_N$) and $b$.

In the presence of layers ($N>1$), $\omega_0$ must also depend on $\epsilon_{N-1}$,
$a_{N-1}$ etc. and therefore condition (\ref{jleq0}) cannot be fulfilled. In fact,
for the same reason, we can assume that also the remaining functions in
conditions (\ref{ANb}) are non-vanishing. Therefore these conditions can be
satisfied by choosing
\begin{align}
A_{lN}^{(1)}& = \frac{\mathcal{N}_l}{j_l(\sqrt{\epsilon_N} kb)},&
A_{lN}^{(2)}& = -\frac{\mathcal{N}_l}{y_l(\sqrt{\epsilon_{\scriptstyle N}} kb)}
\label{ANb1}\\
%
\bar{A}_{lN}^{(1)}& = \frac{\bar{\mathcal{N}}_l}%
{j^{\scriptscriptstyle\mathrm{D}}_l
(\sqrt{\epsilon_N} kb)},&
\bar{A}_{lN}^{(2)}& = -\frac{\bar{\mathcal{N}}_l}%
{y^{\scriptscriptstyle\mathrm{D}}_l(\sqrt{\epsilon_N} kb)}
\qquad
\label{barANb1}
\end{align}
for TE modes (upper line) or TM modes,
where $\mathcal{N}_l$ and $\bar{\mathcal{N}}_l$ are normalization factors.

At the interfaces between dielectrics, the following quantities must be
continuous: the tangential components of the electric field and normal ones of the
magnetic induction and furthermore, the tangential components of the magnetic
induction, due to vanishing of the surface currents at the dielectric surface.
For the TE modes, this leads to the following conditions at $r=a_n$, $n = 1,
\ldots , N-1$:
\begin{equation}\label{contdiel}
\begin{split}
A_{ln}^{(1)} j_l(\rho_n) +
A_{ln}^{(2)} y_l(\rho_n) & =
A_{l\,n+1}^{(1)} j_l(\rho_n^+) +
A_{l\,n+1}^{(2)} y_l(\rho_n^+)\\
A_{ln}^{(1)} j_l^{\scriptscriptstyle\mathrm{D}}(\rho_n) +
A_{ln}^{(2)} y_l^{\scriptscriptstyle\mathrm{D}}(\rho_n) & =
A_{l\,n+1}^{(1)} j_l^{\scriptscriptstyle\mathrm{D}}(\rho_n^+) +
A_{l\,n+1}^{(2)} y_l^{\scriptscriptstyle\mathrm{D}}(\rho_n^+)\\
\rho_n & = \sqrt{\epsilon_n} ka_n\\
\rho_n^+ & = \sqrt{\epsilon_{n+1}} ka_n.
\end{split}
\end{equation}
This can be written in matrix form
\begin{equation}\label{matrform}
\mathbf{M}_n \cdot \mathbf{A}_n = \mathbf{M}_n^+ \cdot \mathbf{A}_{n+1}
\end{equation}
where
\begin{equation}\label{matrdef}
\begin{split}
\mathbf{A}_n &=
\begin{bmatrix}
A_{ln}^{(1)}\\
A_{ln}^{(2)}
\end{bmatrix} \qquad
\mathbf{M}_n = 
\begin{bmatrix}
j_l(\rho_n)&y_l(\rho_n)\\
j_l^{\scriptscriptstyle\mathrm{D}}(\rho_n)&
y_l^{\scriptscriptstyle\mathrm{D}}(\rho_n)
\end{bmatrix}\\
\mathbf{M}_n^+ &= 
\begin{bmatrix}
j_l(\rho_n^+)&y_l(\rho_n^+)\\
j_l^{\scriptscriptstyle\mathrm{D}}(\rho_n^+)&
y_l^{\scriptscriptstyle\mathrm{D}}(\rho_n^+)
\end{bmatrix}.
\end{split}
\end{equation}
The $\mathbf{M}$ matrices are non-singular:
\begin{equation}\label{wronsk}
\begin{vmatrix}
j_l(\rho)&y_l(\rho)\\
j_l^{\scriptscriptstyle\mathrm{D}}(\rho)&
y_l^{\scriptscriptstyle\mathrm{D}}(\rho)
\end{vmatrix} = \frac{\pi}{2}
\begin{vmatrix}
J_{l+\frac{1}{2}}(\rho)&Y_{l+\frac{1}{2}}(\rho)\\
J'_{l+\frac{1}{2}}(\rho)&Y'_{l+\frac{1}{2}}(\rho)
\end{vmatrix} = \frac{1}{\rho} \neq 0
\end{equation}
where last equality follows from the fact that
$J_{l+\frac{1}{2}}(\rho)$ and $Y_{l+\frac{1}{2}}(\rho)$ are solutions
of the Bessel equation
\begin{equation*}
u''(\rho) + \frac{1}{\rho}u'(\rho) + \biggl[1 - \frac{(l+\frac{1}{2})^2}{\rho^2}
\biggr] u(\rho) = 0.
\end{equation*}

Multiplying (\ref{matrform}) by
\begin{equation}\label{inversM}
\mathbf{M}_n^{-1} = \rho_n
\begin{bmatrix}
 y_l^{\scriptscriptstyle\mathrm{D}}(\rho_n)&
-y_l(\rho_n)\\
-j_l^{\scriptscriptstyle\mathrm{D}}(\rho_n)&j_l(\rho_n)
\end{bmatrix}
\end{equation}
we arrive at the recurrence relation
\begin{equation}\label{recur}
\mathbf{A}_n = \mathbf{M}_n^{-1} \cdot \mathbf{M}_n^+ \cdot \mathbf{A}_{n+1}
\qquad n = 1, \ldots , N-1.
\end{equation}
For the $n$th interface between dielectrics, this relation defines the
vector $\mathbf{A}$ at the lower layer in terms of that at the upper one.
Using this relation successively for $n = N-1, N-2, \ldots, 1$, we can express
all $\mathbf{A}_n$ vectors in terms of 
\begin{equation}\label{AN}
\mathbf{A}_N = \mathcal{N}_l
\begin{bmatrix}
\displaystyle{
\frac{1}{j_l(\rho_N)}}\\[2ex]
\displaystyle{
-\frac{1}{y_l(\rho_N)}}
\end{bmatrix} \equiv \mathcal{N}_l \mathbf{a}_N \qquad \rho_N = \sqrt{\epsilon_N} kb
\end{equation}
i.e.,
\begin{eqnarray}
\mathbf{A}_n &=& (\mathbf{M}_n^{-1} \cdot \mathbf{M}_n^+) \cdot
(\mathbf{M}_{n+1}^{-1} \cdot \mathbf{M}_{n+1}^+)\nonumber\\
&& \cdots
(\mathbf{M}_{N-1}^{-1} \cdot \mathbf{M}_{N-1}^+) \cdot \mathbf{A}_N
\equiv \mathcal{N}_l \mathbf{a}_n .\label{Anvec}
\end{eqnarray}

We recall that in the first layer we must satisfy $A_{l1}^{(2)} = 0$, see
(\ref{A1eq0}). In view of this requirement, equations (\ref{contdiel}) for $n=1$
can be written as
\begin{equation}\label{ra1}
\begin{split}
A_{l1}^{(1)} j_l(\rho_1) & -
\mathcal{N}_l \Bigl[a_2^{(1)} j_l(\rho_1^+) +
a_2^{(2)} y_l(\rho_1^+) \Bigr]  = 0\\
A_{l1}^{(1)} j^{\scriptscriptstyle\mathrm{D}}_l(\rho_1) & -
\mathcal{N}_l \Bigl[a_2^{(1)} j^{\scriptscriptstyle\mathrm{D}}_l(\rho_1^+) +
a_2^{(2)} y^{\scriptscriptstyle\mathrm{D}}_l(\rho_1^+)
\Bigr] = 0
\end{split}
\end{equation}
where $\rho_1 = \sqrt{\epsilon_1} ka_1$, $\rho_1^+ = \sqrt{\epsilon_2} ka_1$,
and $a_2^{(1,2)}$ are components of the vector $\mathbf{a}_2 \equiv
\mathbf{A}_2/\mathcal{N}_l$. This vector is defined by (\ref{Anvec}) and
(\ref{AN}) if $N>2$ ($\rho_N = \sqrt{\epsilon_N} kb$):
\begin{eqnarray}
\mathbf{a}_2 &=& (\mathbf{M}_2^{-1} \cdot \mathbf{M}_2^+) \cdot
(\mathbf{M}_3^{-1} \cdot \mathbf{M}_3^+) \cdots
(\mathbf{M}_{N-1}^{-1} \cdot \mathbf{M}_{N-1}^+)\nonumber\\
&&\cdot
\begin{bmatrix}
\displaystyle{
\frac{1}{j_l(\rho_N)}}\\[2ex]
\displaystyle{
-\frac{1}{y_l(\rho_N)}}
\end{bmatrix}.\label{a2vec}
\end{eqnarray}
For $N=2$, $\mathbf{a}_2$ is defined by (\ref{AN}), i.e., is given by the last
factor in (\ref{a2vec}).

The linear and homogeneous set of equations (\ref{ra1}) for $A_{l1}^{(1)}$
and $\mathcal{N}_l$ will have non-zero solutions if and only if its
determinant vanishes,
\begin{eqnarray}
&&j_l(\rho_1) \Bigl[
a_2^{(1)} j^{\scriptscriptstyle\mathrm{D}}_l(\rho_1^+) +
a_2^{(2)} y^{\scriptscriptstyle\mathrm{D}}_l(\rho_1^+)
\Bigr]\nonumber\\
&&- j^{\scriptscriptstyle\mathrm{D}}_l(\rho_1) \Bigl[
a_2^{(1)} j_l(\rho_1^+) +
a_2^{(2)} y_l(\rho_1^+) \Bigr] = 0.\label{compat}
\end{eqnarray}
If this condition is fulfilled, $A_{l1}^{(1)}$ is given by either of equations
(\ref{ra1}), which are equivalent. Like all remaining coefficients
$A_{ln}^{(1)}$ and $A_{ln}^{(2)}$, $n=2, \ldots ,N$, also $A_{l1}^{(1)}$
will be proportional to the normalization factor $\mathcal{N}_l$, see
(\ref{Anvec}) and (\ref{ra1}).

If there are only two layers ($N=2$), $a_2^{(1)}$ and $a_2^{(2)}$
in (\ref{ra1}) and (\ref{compat}) are given by (\ref{AN}) and the
resonant frequency $\omega_0$ defined by (\ref{compat}) can be found from
\begin{eqnarray}
&&j_l(\sqrt{\epsilon_1} ka_1) \biggl[
\frac{j^{\scriptscriptstyle\mathrm{D}}_l(\sqrt{\epsilon_2} ka_1)}{
j_l(\sqrt{\epsilon_2} kb)} -
\frac{y^{\scriptscriptstyle\mathrm{D}}_l(\sqrt{\epsilon_2} ka_1)}{
y_l(\sqrt{\epsilon_2} kb)}
\biggr]\nonumber\\
&&- j^{\scriptscriptstyle\mathrm{D}}_l(\sqrt{\epsilon_1} ka_1) \biggl[
\frac{j_l(\sqrt{\epsilon_2} ka_1)}{
j_l(\sqrt{\epsilon_2} kb)} -
\frac{y_l(\sqrt{\epsilon_2} ka_1)}{
y_l(\sqrt{\epsilon_2} kb)} \biggr] = 0.\nonumber\\
\label{compN2}
\end{eqnarray}
and
\begin{equation}
A_{l1}^{(1)} = \mathcal{N}_l \frac{1}{j_l(\sqrt{\epsilon_1} ka_1)} \biggl[
\frac{j_l(\sqrt{\epsilon_2} ka_1)}{
j_l(\sqrt{\epsilon_2} kb)} -
\frac{y_l(\sqrt{\epsilon_2} ka_1)}{
y_l(\sqrt{\epsilon_2} kb)} \biggr].
\label{coefN2}
\end{equation}

By replacing in (\ref{contdiel})--(\ref{coefN2})
\begin{equation}\label{reps1}
\begin{split}
&A_{lm}^{(1,2)} \to \bar{A}_{lm}^{(1,2)}, \qquad
j^{\scriptscriptstyle\mathrm{D}}_l(\sqrt{\epsilon_m} ka) \to
j^{\scriptscriptstyle\mathrm{D}}_l(\sqrt{\epsilon_m} ka)/\epsilon_m,\\
&y^{\scriptscriptstyle\mathrm{D}}_l(\sqrt{\epsilon_m} ka) \to
y^{\scriptscriptstyle\mathrm{D}}_l(\sqrt{\epsilon_m} ka)/\epsilon_m
\end{split}
\end{equation}
for any $m$ and $a$, we obtain the corresponding equations for the TM modes.

Any standard software like \textit{Mathematica} or \textit{Maple} can be used
to solve the non-linear equation (\ref{compat}) or (\ref{compN2}) defining the
resonant frequency $\omega_0 \equiv \omega^l$, along with the pertinent linear
algebra for $N>2$. We did it for $N=2$, see the
following section, and also for $N=3$, by using \textit{Mathematica}.

Note that for the TM modes, where $\bar{\mathcal{B}}_{\mathrm{t}}(r)$ in
(\ref{barElmr}) is continuous at each dielectric interface,
$\bar{\mathcal{E}}_{\mathrm{r}}(r)$ will have jumps, due to discontinuities in
$\epsilon$. However, the radial component of the electric displacement
$\bar{\mathcal{D}}_{\mathrm{r}}(r) \equiv \epsilon \bar{\mathcal{E}}_{\mathrm{r}}(r)$
will be continuous. This will also be true of the TE modes where the radial
displacement is identically zero. These facts imply the vanishing of surface
charges at each dielectric interface. And this in turn means that the multi-layer
dielectric structure resembles (and can approximate) a smooth dielectric with some
permittivity profile $\epsilon(r)$, in spite of jumps in $\epsilon$.

It was pointed out to us by Paul Martin of SIAM, that our matrix equation
(\ref{matrform}), which can be used to relate the EM fields of a given mode for
two layers of a stratified sphere, is not new. It was probably first used by
A. Moroz \cite{moroz} when calculating forced oscillations in such a sphere but
without a conducting wall, induced by an oscillating electric dipole. In this
application, the frequency $\omega$ is arbitrary.

\section{A spherical conductive cavity with a dielectric sphere}

The general theory given in the previous section will now be illustrated by
calculations  pertinent to the TE modes in a spherical cavity with a
dielectric sphere of radius $a$ and dielectric permittivity $\epsilon$, i.e.,
for $N=2$, $a_1 \equiv a$, $\epsilon_1 \equiv \epsilon$, and $\epsilon_2 = 1$.

Fields in such a system will be described by equations (\ref{Elm}) -- (\ref{Blmr}) both
in the sphere and the surrounding vacuum. In view of (\ref{A1eq0}) and (\ref{ANb1})
their radial profiles will be given by
\begin{equation}
\label{e1}
\mathcal{E}_{\text{t}}(r) = \mathcal{N}_l \times \left\{
\begin{array}{ll}
\mathcal{A}_l j_l(\sqrt{\epsilon} kr) &  \text{if}\quad 0 \le  r \le a\\[1ex]
\displaystyle{
\frac{j_l(kr)}{j_l(kb)} - \frac{y_l(kr)}{y_l( kb)}} &
\text{if} \quad a \le r \le b
\end{array} \right.
\end{equation}
\begin{equation}
\label{ b1}
\mathcal{B}_{\text{t}}(r) = -\frac{i\mathcal{N}_l}{kr} \times \left\{
\begin{array}{ll}
\mathcal{A}_l \, j^{\scriptscriptstyle\text{D}}_l(\sqrt{\epsilon} kr)& \text{if}
\quad  0
\le  r \le a\\[1ex]
\displaystyle{
\frac{j^{\scriptscriptstyle\text{D}}_l(kr)}{j_l(kb)} -
\frac{y^{\scriptscriptstyle\text{D}}_l(kr)}{y_l(kb)}}&
\text{if} \quad  a \le r \le b .
\end{array}\right.
\end{equation}
Replacing $\epsilon_1 \to \epsilon$, $\epsilon_2 \to 1$ and $A_{l1}^{(1)} \to
\mathcal{A}_l$ in (\ref{compN2}) and (\ref{coefN2}), we obtain equations
defining the resonant frequency $\omega^l$ and the amplitude coefficient
$\mathcal{A}_l$.

We verified that for $\omega = \omega^l$, the average energies associated with
the electric and the magnetic field in the cavity are equal:
\begin{eqnarray}
&&\int_V \epsilon' | \mathbf{E}_{lm} (\mathbf{r},\omega^l) |^2 \, \text{d}v\nonumber\\
&&= \int_V \bigl( | \mathbf{B}_{lm\, \text{t}}(\mathbf{r},\omega^l) |^2
+ | \mathbf{B}_{lm\, \text{r}} (\mathbf{r}, \omega^l) |^2 \bigr) \, \text{d}v.
\label{energies}
\end{eqnarray}
(This was a check on the correctness of our formulas and
accuracy of calculations.) The normalization constant $\mathcal{N}_l$ was chosen so
as to satisfy:
\begin{eqnarray}
&&\frac{1}{2} \biggl( \int_0^a \epsilon' \lvert\mathcal{E}_{\text{t}}(r)\rvert^2
r^2 \text{d}r +  \int_a^b \lvert\mathcal{E}_{\text{t}}(r)\rvert^2 r^2 \text{d}r\nonumber\\
&&+ \int_0^b \Bigl(\lvert\mathcal{B}_{\text{t}}(r)\rvert^2 +
\lvert\mathcal{B}_{\text{r}}(r)\rvert^2 \Bigr) r^2 \text{d}r \biggr) = 1.
\label{norm}
\end{eqnarray}
(The corresponding average energy associated with the electric and
the magnetic field over our cavity is $1/(8\pi)$ erg.)

\begin{figure}[h]
\centering\includegraphics[scale=.5]{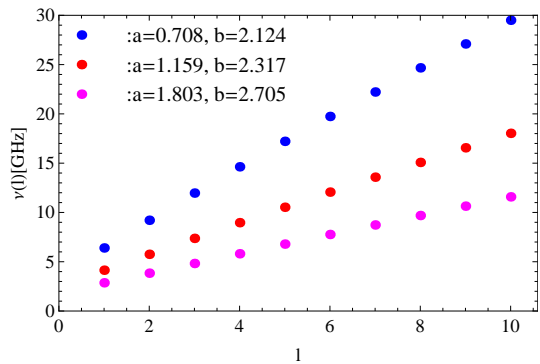}%
\caption{The resonant frequencies $\nu(l) = \omega^l/(2\pi)$ versus $l$
for three spherical cavities with dielectric spheres ($a$ and $b$ in cm).
Note that $\nu$ is $m$ independent (due to degeneracy).}\label{nuofl}
\end{figure}

\begin{figure}
\centering\includegraphics[scale=.5]{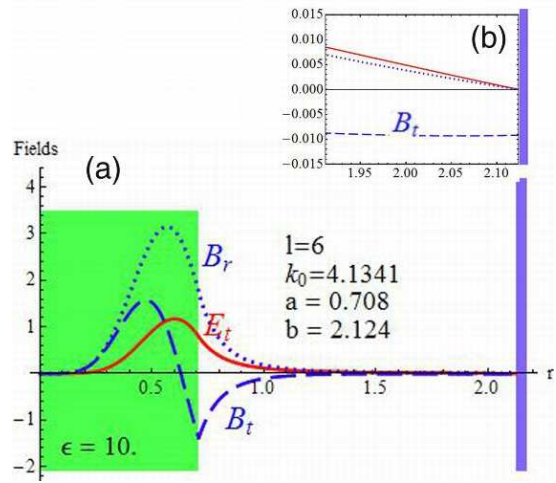} \caption{(a) Field
radial functions for a spherical cavity with dielectric sphere and
perfect metallic wall (Gaussian units, $k_0 = \omega^l/c$, $a$ and
$b$ in cm). The complete fields are given by (\ref{Elm}) -- (\ref{Br}).
Note that $\mathcal{E}_{\text{t}}$ and $\mathcal{B}_{\text{r}}$
vanish at the wall, whereas $\mathcal{B}_{\text{t}}$ is non-zero but
small, see (b).} \label{fields}
\end{figure}

In \textbf{Figure \ref{nuofl}}, $\nu(l) = \omega^l/(2\pi)$ as a
function of $l$ is presented for three spherical cavities with dielectric
spheres.
Note that $\nu$ is $m$ independent (due to degeneracy). This fact is one of the
reasons why the spherical resonator shown in \textbf{Figure \ref{accunits}} cannot
be used in the final project of a real accelerator even though it is very convenient
for a general analysis. The degeneration in question can be broken e.g., by
replacing the spherical resonator by an ellipsoidal one, or shifting the center
of the dielectric sphere. Another possibility could be to use an
anisotropic dielectric. In any case, however, a separate numerical analysis would
be necessary.

In \textbf{Figure \ref{fields}} we give an example of radial functions in the
equations describing fields in our spherical cavity with a dielectric sphere,
(\ref{Elm}), (\ref{Blmt}) and (\ref{Blmr}). Large values of these functions in a
vicinity of the dielectric boundary can be observed.

\subsection{The motion of relativistic electrons}

The trajectory $\mathbf{r}(t)$ of a relativistic electron crossing  the spherical
cavity shown in \textbf{Figure 1} can be parametrized by the
electron's closest approach $\mathbf{r}_0 = \mathbf{r}(t_0)$  and electron
velocity $c\vec{\beta}$, $|\vec{\beta}| = 1$:
\begin{equation}
\label{eqtra}
\mathbf{r}(t) = \mathbf{r}_0 + c\vec{\beta} (t - t_0).
\end{equation}

The origin of the Cartesian coordinate system $(x,y,z)$ was chosen at
the center of the dielectric sphere, and the electron moving along the $x$
axis was passing just above the dielectric sphere as shown in \textbf{Figure 1}. We
chose
\begin{equation}
\mathbf{r}_0 = 1.01 a ( 0,\sin\theta_0,\cos\theta_0 ), \qquad \vec{\beta} =
(1,0,0).\label{R0}
\end{equation}

\begin{figure}[h]
\centering\includegraphics[scale= 0.55]{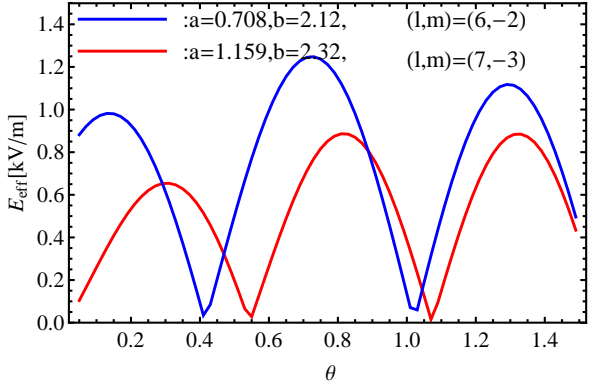} \caption{Effective
accelerating field as a function of $\theta_0$ for two spherical cavities
($a$ and $b$ in cm).}\label{Eefvthet}
\end{figure}

\begin{figure}[b]
\centering\includegraphics[scale=.55]{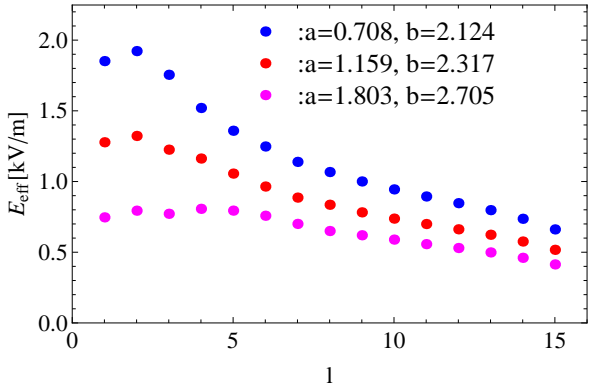} \caption{Effective accelerating
fields versus $l$ for three spherical cavities with the assumed normalization of
the RF field, see (\ref{norm}) ($a$ and $b$ in cm). For each value of $l$,
these values were optimized with respect to $m$ as well as $\theta_0$,
see \textbf{Figure \ref{Eefvthet}}.}\label{Eefvl}
\end{figure}

\begin{figure}[h]
\centering\includegraphics[scale=.55]{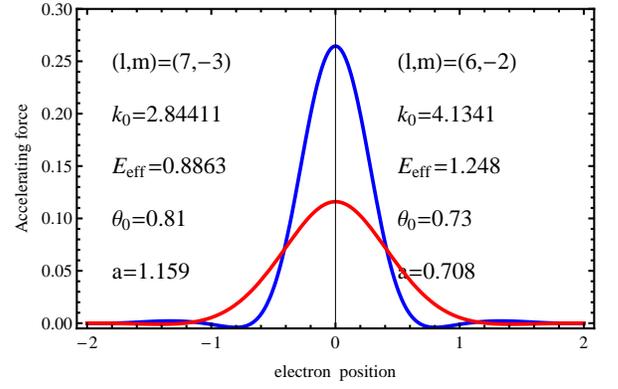}
\caption{The accelerating force (in units of $4.8032 \times 10^{-10}$ dyne)
acting on an electron as it moves through the cavity shown in
\textbf{Figure \ref{accunits}} ($k_0 = \omega^l/c$, $E_{\text{eff}}$ in kV/m, and
dimensions in cm).}\label{accf}
\end{figure}

The effective accelerating field felt by the electron as it passes through the
cavity, $E_{\text{eff}}$, is equal to the real part of ($c \, d t = dx$)
\begin{equation}
\bar {E}(l,m,\theta) \equiv |\bar{E}| e^{i \varphi} =
\frac{c}{d} \int_{t_0 - \frac{d}{2c}}^{t_0 + \frac{d}{2c}}
E_x \bigl(\mathbf{r}(t),\omega^l \bigr)
e^{-i\omega^l t} \, \text{d} t,
\label{W}
\end{equation}
where $d = 2 \sqrt{b^2 - (1.01 a)^2}$ is the electron trajectory segment within
the cavity, $E_x$ is the $x$ component of the electric field $\mathbf{E}^{lm}
(\mathbf{r},\omega^l)$ given by (\ref{Elm}) and $\mathbf{r}(t)$ is given by
(\ref{eqtra}). Thus
\begin{equation}
E_{\text{eff}} = |\bar {E}| \cos \varphi, \qquad \varphi =
f(l,m,\theta) - \omega^l t_0.\label{Eeff}
\end{equation}
Maximal acceleration is obtained ($E_{\mathrm{eff}} =
|\bar {E}|$) if $t_0$ is chosen so that the accelerating phase $\varphi = 0$.
With this choice, the relativistic electron is never decelerated within the
spherical cavity, see \textbf{Figure \ref{accf}}, where two examples are given.
Typical results for $E_{\mathrm{eff}}|_{\varphi = 0}$ obtained with our normalization
(\ref{norm}) are shown in \textbf{Figures \ref{Eefvthet}} and \textbf{\ref{Eefvl}}.


The electromagnetic field given by the real parts of (\ref{Elm}), (\ref{Blmt})
and (\ref{Blmr}) is strongly non-uniform. Therefore one should check how much the
relativistic electron will deflect from the assumed trajectory $\mathbf{r}(t)$
given by (\ref{eqtra}), due to interaction with this field. A nice feature of our
model is that the field in question is described analytically by
(\ref{Elm})--(\ref{Blmr}) so that the pertinent equations of
the transversal motion can easily be integrated numerically.

\begin{figure}[t]
\centering\includegraphics[scale = .5]{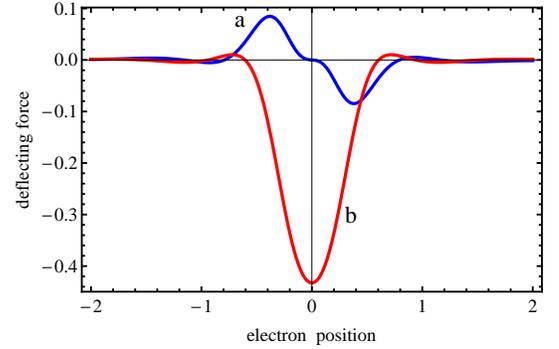} \caption{The deflecting force
coordinate $F_{\bar{y}}(\bar{x})$ for the accelerating phase $\varphi = 0$ (a)
and $\varphi = \pi/2$ (b), see \textbf{Figure \ref{accf}} for units and parameters
(right column).}\label{df1}
\end{figure}
\begin{figure}[h]
\centering\includegraphics[scale = .5]{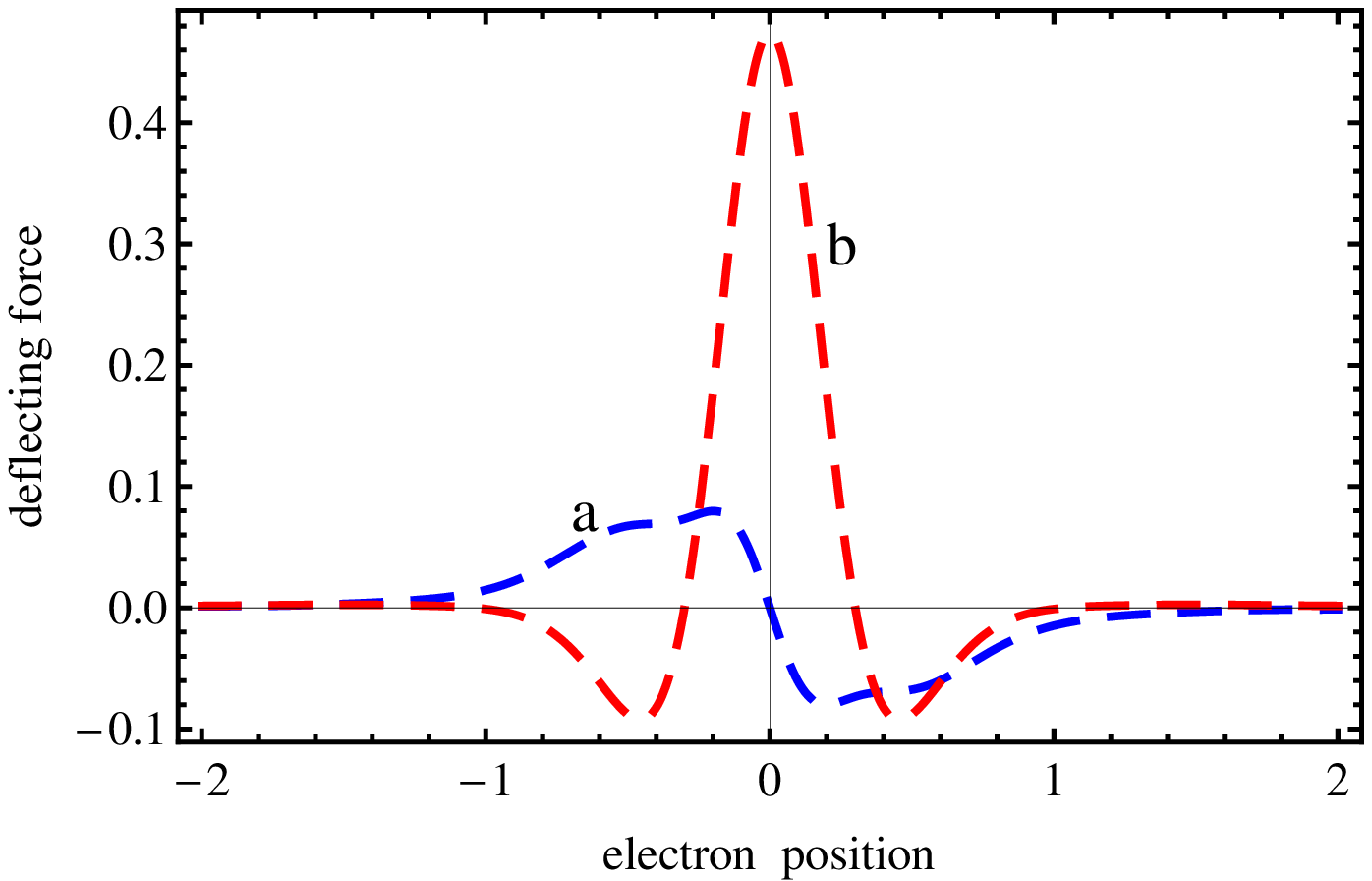} \caption{The deflecting force
coordinate $F_{\bar{z}}(\bar{x})$ for the accelerating phase $\varphi = 0$ (a)
and $\varphi = \pi/2$ (b), see \textbf{Figure \ref{accf}} for units and parameters (right column).}\label{df2}
\end{figure}
\begin{figure}[h]
\centering\includegraphics[scale = .5]{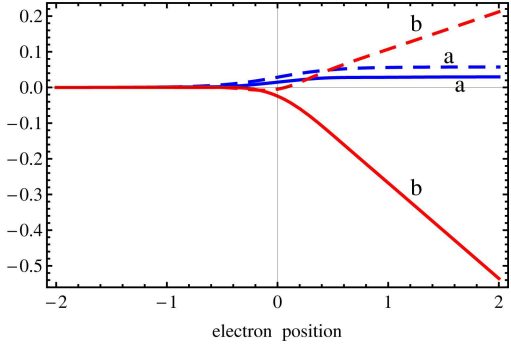} \caption{Conveniently normalized
transversal displacements $\bar{y}$ (solid curves) and $\bar{z}$ (dashed curves) as
functions of $\bar{x}$ for the accelerating phase $\varphi = 0$ (a) and $\varphi =
\pi/2$ (b), see \textbf{Figure \ref{accf}} for units and parameters (right column).}
\label{ndisp}
\end{figure}

In a real accelerator, where we are dealing with an electron beam of finite cross
section, the electromagnetic fields $\mathbf{E}\bigl(\mathbf{r}(t),t\bigr)$ and
$\mathbf{B}\bigl(\mathbf{r}(t),t\bigr)$ acting on each electron will be
superpositions of the external fields and the fields due to the electron charge
and current. However, in the lowest approximation (and particularly for not too
large beam densities) the latter fields can be neglected. Furthermore, if as
in our case, the transversal deflections are small, the deflecting fields can be
calculated on the unperturbed trajectory given by (\ref{eqtra}). It will also be
assumed that the electron mass $m_{\text{e}} = m_{\text{e}0} \, \gamma$,
$\gamma \gg 1$, is time independent within the spherical cavity. With these
approximations, and within the Cartesian coordinate system
$(\bar{x},\bar{y},\bar{z})$ with its center at $\mathbf{r}_0$, the $\bar{x}$ axis
along the unperturbed trajectory, and the $\bar{y}$ axis along $\mathbf{r}_0$,
the electron's transversal motion will be described by
%
\begin{eqnarray}
m_{\text{e}} \frac{\text{d}v_{\bar{y}}}{\text{d}\bar{t}} &=& -e \bigl[ E_{\bar{y}} -
B_{\bar{z}} \bigr], \qquad v_{\bar{y}} = \frac{\text{d}\bar{y}}{\text{d}\bar{t}}\label{trma}\\
m_{\text{e}} \frac{\text{d}v_{\bar{z}}}{\text{d}\bar{t}} &=& -e \bigl[ E_{\bar{z}} +
B_{\bar{y}} \bigr], \qquad v_{\bar{z}} = \frac{\text{d}\bar{z}}{\text{d}\bar{t}}
\label{trmb}
\end{eqnarray}
%
where $\bar{t} = t - t_0$, the field components $E_{\bar{y}}(\bar{x}, \bar{t})$,
$B_{\bar{z}}(\bar{x}, \bar{t})$, etc. are given by the real parts of
Eqs.~(\ref{Elm}), (\ref{Blmt}) and (\ref{Blmr}), taken at $\bar{x} = c \bar{t}$,
and $\bar{y} = \bar{z} = 0$.
Integrating these equations with zero initial conditions we end up with
($d\bar{x} = c d\bar{t}$)
\begin{eqnarray}
v_{\bar{y}}(\bar{x}) &=& - \frac{e}{m_{\text{e}} c} \int_{-d/2}^{\bar{x}}
F_{\bar{y}}(\bar{x}')
\, \text{d}\bar{x}'\label{resa}\\
v_{\bar{z}}(\bar{x}) &=& - \frac{e}{m_{\text{e}} c} \int_{-d/2}^{\bar{x}}
F_{\bar{z}}(\bar{x}')
\, \text{d}\bar{x}'\label{resb}
\end{eqnarray}
\begin{equation}
\bar{y}(\bar{x}) = \frac{1}{c} \int_{-d/2}^{\bar{x}} v_{\bar{y}} \,
\text{d}\bar{x}', \qquad \bar{z}(\bar{x}) = \frac{1}{c} \int_{-d/2}^{\bar{x}}
v_{\bar{z}} \, \text{d}\bar{x}'\label{resc}
\end{equation}
where $F_{\bar{y}} = E_{\bar{y}} - B_{\bar{z}}$, and $F_{\bar{z}} =
E_{\bar{z}} + B_{\bar{y}}$.

In \textbf{Figures \ref{df1}} and \textbf{\ref{df2}} we give an example of the
coordinates $F_{\bar{y}}(\bar{x})$ and $F_{\bar{z}}(\bar{x})$ of the deflecting
force. They correspond to $a = 0.708$ cm, $b = 2.124$ cm, and $(l,\ m) =
(6,\ -2)$, for which the the accelerating field will be our reference value
\begin{equation}
E_{\text{eff}}\Bigr|_{\varphi = 0} \equiv E_{\text{eff\:ref}} = 1.2366 \ 
\text{kV/m}.\label{Eref}
\end{equation}
It can be seen that if the accelerating phase $\varphi = 0$, both
$F_{\bar{y}}(\bar{x})$ and $F_{\bar{z}}(\bar{x})$ are odd functions. Therefore
in this case of maximal acceleration, there will be no transversal velocity
increments over the cavity, $v_{\bar{y}}(d/2) = v_{\bar{z}}(d/2) = 0$.
At the same time the velocity components $v_{\bar{y}}(\bar{x})$ and
$v_{\bar{z}}(\bar{x})$ in (\ref{resc}) will be even functions of $\bar{x}$
tending to zero as $\bar{x} \to d/2$.
Hence the transversal deflections $\bar{y}(\bar{x})$ and $\bar{z}(\bar{x})$
will be increasing functions tending to constants as $\bar{x} \to d/2$, see
\textbf{Figure \ref{ndisp}(a)}.

For the worst case of accelerating phase ($\varphi = \pi/2$) for which
$E_{\text{eff}} = 0$, $v_{\bar{y}}(\bar{x})$ and $v_{\bar{z}}(\bar{x})$ will
be increasing functions soon reaching their limiting values $v_{\bar{y}}(d/2)$
and  $v_{\bar{z}}(d/2)$ for $\bar{x} > 0$. The corresponding transversal motions
$\bar{y}(\bar{x})$ and $\bar{z}(\bar{x})$ will soon become uniform for
$\bar{x} > 0$, leading to much larger deflections at $\bar{x} = d/2$,
see \textbf{Figure \ref{ndisp}(b)}.

The actual transversal deflections per cavity in an accelerator involving our
spherical cavities can be obtained by multiplying the normalized values given in
\textbf{Figure \ref{ndisp}} by the factor
\begin{equation}
\alpha = \frac{e}{m_{\text{e}}} \frac{E_{\text{eff}}}{E_{\text{eff\:ref}}} \ 
\text{cm} = 5.9 \times 10^{-4} \frac{E_{\text{eff}}/E_{\text{eff\:ref}}}{\gamma}
\ \text{cm}\label{alpha}
\end{equation}
where $E_{\text{eff\:ref}}$ is given by (\ref{Eref}), $E_{\text{eff}}$ is
the assumed value of the effective accelerating field, and
$\gamma = m_{\text{e}}/m_{\text{e}0}$. For large values of $E_{\text{eff}}$,
small $\alpha$ requires $\gamma$ to be sufficiently large. Assuming that
$\bar{z}_{\text{max}} \, \bigl(= \bar{z}(d/2) > \bar{y}(d/2)$, see \textbf{Figure
\ref{ndisp} (a)}$\bigr)$ is not larger than $p \, \%$ of the spacing between the dielectric
sphere and the electron trajectory, $0.01 a$, $a = 0.708$ cm, the required minimal
electron energy is given by
\begin{equation}
m_{\text{e}}\,c^2 \: [\text{GeV}] = \frac{20.69}{p} \ \frac{E_{\text{eff}} \:
[\text{MV/m}]}{100}.\label{mecs}
\end{equation}
Thus, if we assume that $E_{\text{eff}} = 100$ MV/m, the transversal
displacements will be smaller than $1 \%$ of the spacing in question, if
the electron energy $m_{\text{e}}\,c^2 \geq 21$ GeV, i.e., for typical output
energies from SLAC.  Whether the real dielectric can withstand this value of
$E_{\mathrm{eff}}$ is another question beyond the scope of this paper. More
comments will be given later on.

\subsection{Quality factors}

An important parameter of any linear accelerator is the quality $Q$ of its
resonant cavities:
\begin{equation}
Q = \omega_0 \frac{U}{P} \equiv 2 \pi \frac{U}{T_0 P}\label{Qdef}
\end{equation}
where $\omega_0$ is the resonant angular frequency of the ideal cavity
($\sigma = \infty, \epsilon'' = 0$), $T_0$ is the corresponding resonant period,
$U$ is the time-averaged energy stored in the cavity ($\mu = 1$)
\begin{equation}
U = \frac{1}{16\pi} \int_{V} \bigl( \epsilon' \lvert\mathbf{E}\rvert^2 +
\lvert\mathbf{H}\rvert^2 \bigr) \, \text{d}v\label{U}
\end{equation}
and $P$ is time-averaged cavity power loss.

The power loss caused by the skin current in the metallic wall bounded by
the surface $S$ is given by
\begin{equation}
P_{\mathrm{met}} = \alpha \int_{S} |\mathbf{H}|^2 \, \text{d} s
\end{equation}
where
\begin{equation}
\alpha = \frac{c}{8(2\pi)^{3/2}} \sqrt{\frac{\omega_0}{\sigma}} \equiv
\frac{c^2}{32 \pi^2 \sigma \delta},
\qquad \delta = \frac{c}{\sqrt{2\pi\sigma\omega_0}}
\end{equation}
$\delta$ is the skin depth, $\sigma$ is conductivity of the wall, and the
magnetic field intensity $\mathbf{H}$ refers to the ideal cavity, i.e.,
its normal component is vanishing ($\mathbf{H} = \mathbf{H}_{\mathrm{t}}$).

The quality of the cavity related to losses in the metallic wall is thus given by
\begin{equation}
Q_{\text{met}} = \omega_0 \frac{U}{P_{\text{met}}}.
\end{equation}

Using the fact that at resonance, the averaged energies stored in the electric
and magnetic fields are equal, see (\ref{energies}), we end up with
($\mathbf{H} = \mathbf{B}$):
\begin{equation}
\label{Q}
Q_{\text{met}} = \frac{2}{\delta} \, \frac{\int_{V} \lvert\mathbf{B}\rvert^2
\, \text{d}v}{\int_{S} \lvert\mathbf{B}\rvert^2 \, \text{d}s}.
\end{equation}
This formula is quite general, and in particular can also be used for a
traditional cylindrical cavity of radius $R_{\text{c}}$ and height $h$. In
that case, the cylindrically symmetric $n=0$ TM mode used for acceleration
is given by
\begin{equation}
E_x(\rho,t) = \mathcal{N} J_0 (k_0 \rho) \exp(- i \omega_0 t)
\label{Ex}
\end{equation}
\begin{equation}
B_{\varphi}(\rho,t) = {i \mathcal{N}} J^{\prime}_0( k_0 \rho )\exp(- i
\omega_0 t )\label{Bphi}
\end{equation}
where $k_0 = \omega_0/c$,
$J_0$ is a Bessel function, and $\rho$ and $\varphi$ are cylindrical coordinates
(cylindrical axis along $x$).

\begin{figure}[h]
\centering\includegraphics[scale=.5]{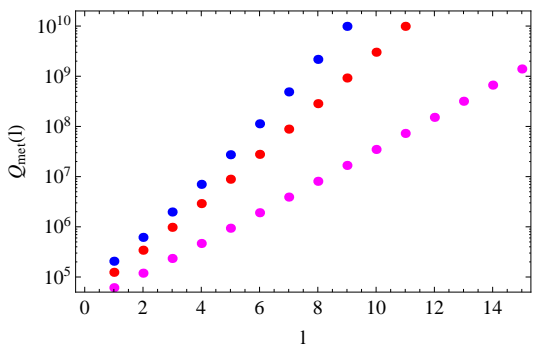} \caption{The quality
$Q_{\text{met}}$ versus $l$ for three spherical cavities with ideal dielectric
spheres (see \textbf{Figure \ref{Eefvl}}).}\label{Qmet}
\end{figure}

The vanishing of $E_x$ on an ideally conducting cylindrical wall requires that
$(\omega_0/c) R_{\text{c}} = 2.405$ (the smallest zero of $J_0$) which defines
the angular resonant frequency in terms of $R_{\text{c}}$.
Equations (\ref{Bphi}) and (\ref{Q}) lead to the well known formula for the quality
of the cylindrical pill box cavity
\begin{equation}
\label{Qpb}
Q_{\text{c}} = 2.405 \sqrt{\frac{2\pi\sigma}{\omega_0}} \, \frac{1}{1 +
R_{\text{c}}/h}.
\end{equation}
For the  SLAC  pill box cavity shown in \textbf{Figure \ref{accunits}} ($R_{\text{c}} =
h = d = 4$ cm, and $\sigma = 5.294 \times 10^{17}$ s$^{-1}$ for copper wall in
room-temperature), this formula leads to $Q_{\text{c}} = 1.633 \times 10^4$.
The corresponding values for spherical cavities with ideal dielectric spheres
and the same values of $d$ and $\sigma$ reach much larger values, see
\textbf{Figure \ref{Qmet}}.

In the presence of the dielectric sphere, one is also dealing with
losses due to an imperfect dielectric specified by $\epsilon'' = \text{Im}
\, \epsilon$. The non-vanishing value of $\epsilon''$ leads to $\text{Im}
\, k \neq 0$, for $k$ defined by (\ref{compat}). In view of Eq.~(\ref{kvom}) this
implies a complex value of $\omega = \omega' + i \omega''$.

For the fields given by (\ref{Elm}) -- (\ref{Blmr}), we obtain
\[
U(t) = U(t = 0) e^{2 \omega'' t}
\]
where $\omega'' < 0$ for the energy $U$ being dissipated rather than generated.
Using this result we find for the power losses in the dielectric:
\[
P_{\text{diel}} = - \frac{\text{d}U}{\text{d}t} = - 2\omega'' U.
\]
In view of (\ref{Qdef}), the corresponding quality will thus be given by
\begin{equation}
\label{Qd}
Q_{\text{diel}} = - \frac{\omega_0}{2 \text{Im} \, \omega}.
\end{equation}
This value is of the order of $(\tan \delta)^{-1} \equiv \epsilon'/\epsilon''$.
It is approximately $l$ independent.

In our calculations we took $\epsilon' = 10$ and $\epsilon'' = 10^{-6}$.
Dielectrics with such ultra small losses were investigated in \cite{krupka}.

\begin{figure}[h]
\centering\includegraphics[scale=.5]{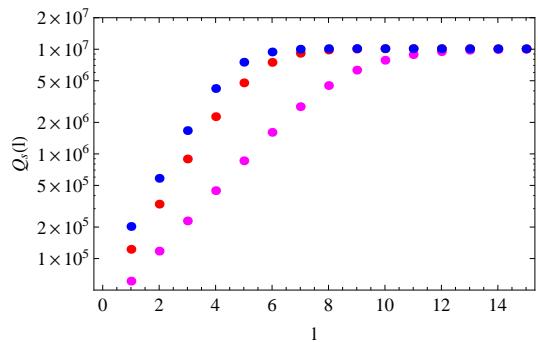}%
\caption{The quality $Q_{\text{s}}$
versus $l$ for three spherical cavities with dielectric spheres
(see \textbf{Figure \ref{Eefvl}}).}\label{Qs}
\end{figure}

The total power loss in the spherical cavity encasing the dielectric sphere $P_{\text{s}}$
is due to the power loss in the metallic wall and that in the dielectric sphere:
\begin{equation}
P_{\text{s}} = P_{\text{met}} + P_{\text{diel}}.\label{Ps}
\end{equation}
Dividing both sides of this relation by $\omega_0 \, U$ and using (\ref{Qdef}) we obtain
\begin{equation}
\frac{1}{Q_{\text{s}}} = \frac{1}{Q_{\text{met}}} + \frac{1}{Q_{\text{diel}}}.\label{Qseq}
\end{equation}
where $Q_{\text{s}}$ is the total Q-factor of the spherical cavity.
Values of $Q_{\text{s}}$ versus $l$ for three spherical cavities with dielectric
spheres and $d=4$ cm are shown in \textbf{Figure~\ref{Qs}}. They are about three
orders of magnitude larger than $Q_{\text{c}} = 1.633 \times 10^4$.

In a real accelerator, openings in the metallic wall are necessary for free
penetration of the cavity by the electron beam, and to enable coupling between
neighbouring cavities. This will lower the quality $Q_{\mathrm{met}}$ but should
have little effect on the total quality of the spherical cavity $Q_{\mathrm{s}}$.
The latter is defined by losses in the dielectric, see (\ref{Qseq}) where
$Q_{\mathrm{diel}} \ll Q_{\mathrm{met}}$. The resonant frequency should not
be drastically changed either, as the EM fields at the iris ($r = b$) are
very small fractions of their maxima, see \textbf{Figure \ref{fields}}.

\subsection{Discussion and summary}

When comparing the effective accelerating fields in the traditional pill box
cavity with that in our spherical cavities with dielectric spheres, we first
assume that $U$, the time-averaged energy stored in the cavity, see
(\ref{U}), is the same in both situations. Therefore the normalization factor
$\mathcal{N}$ in (\ref{Ex}) and (\ref{Bphi}) will first be chosen so that
\begin{equation}
2 \pi h \frac{1}{2} \int_0^{R_{\text{c}}} \rho \, \text{d}\rho  \Bigl[ \lvert
E_x(\rho,t) \rvert^2 + \lvert B_{\varphi}(\rho,t)\rvert^2 \Bigr] = 1
\end{equation}
see (\ref{norm}) ($U = 1/(8\pi)$ erg).

The complex effective accelerator field $\bar{E}$ for the cylindrical resonator
shown in \textbf{Figure \ref{accunits}} ($R_{\text{c}} = h = d$) is given by the
right hand side of (\ref{W}) in which $E_x$ is defined by (\ref{Ex}) with
$\rho = 0$, and $\omega^l = \omega_0$. The result is

\begin{equation}
\bar{E}_{\text{c}} = \mathcal{N} \: \frac{\sin\alpha}{\alpha} \,
e^{-i\omega_0 t_0}, \qquad \alpha = \frac{\omega_0 d}{2c}
\end{equation}
where $t_0$ is the time at which the electron passes the center of the cavity.

We now denote by $E_{\mathrm{eff_s}}$ and $E_{\mathrm{eff_c}}$ the maximal
effective accelerating fields (equal to $|\bar{E}|$) for the spherical cavity
with a dielectric sphere and the traditional cylindrical cavity, for any values
of the average energies in the cavities, $U_{\mathrm{s}}$ and $U_{\mathrm{c}}$.
In view of the fact that $E_x$ in (\ref{Ex}) is proportional to $\sqrt{U}$
we can write, using the definition (\ref{Qdef}) of $Q$,
\begin{equation}
\frac{E_{\mathrm{eff_s}}}{E_{\mathrm{eff_c}}} =
\sqrt{\frac{P_{\text{s}}}{P_{\text{c}}}} G(l)\label{DeltaE}
\end{equation}
where $G(l)$, the ``gain factor'', is given by
\begin{equation}\label{Geq}
G(l) = \left.\frac{E_{\mathrm{eff_s}}}{E_{\mathrm{eff_c}}}
\right|_{U_{\text{s}} = U_{\text{c}}}\sqrt{
\frac{Q_{\text{s}}}{Q_{\text{c}}} \ \frac{\nu_{\text{c}}}{\nu_{\text{s}}}}.
\end{equation}
Here $\nu_{\mathrm{s}}$ and $\nu_{\mathrm{c}}$ are the resonant frequencies of
the spherical and the cylindrical cavities $\bigl(\nu = \omega_0/(2\pi)\bigr)$
and $P_{\mathrm{s}}$ and $P_{\mathrm{c}}$ are the corresponding power losses.
They are equal to the powers that must be supplied from external sources to
sustain the oscillations. They should be as large as possible to avoid breakdown
in the dielectric or at the metallic wall. Further research is necessary to give
an estimate of the ratio $P_{\text{s}}/P_{\text{c}}$. We can only hope that it
is not smaller than unity.

\begin{figure}[h]
\centering\includegraphics[scale = .5]{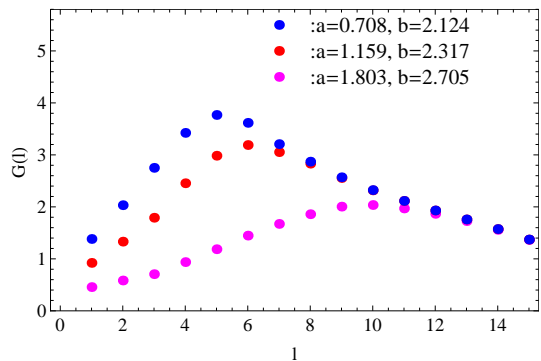} \caption{Gain
factor $G(l)$ in (\ref{DeltaE}) and (\ref{Geq}) for three
spherical cavities ($a$ and $b$ in cm).}\label{Gofl}
\end{figure}

For our typical  SLAC  pill box cavity shown in \textbf{Figure \ref{accunits}}
($R_{\mathrm{c}} = h = d = 4$ cm) we obtain $\nu_{\mathrm{c}} = 2.87$ GHz and
$E_{\mathrm{eff_c}} \bigr|_{U_{\mathrm{c}} = 1/(8\pi) \mathrm{\ erg}} =
3.16$ kV/m, to be used in (\ref{Geq}). The resulting values of the gain factor
are shown in \textbf{Figure \ref{Gofl}}.

The results of our calculation are shown in \textbf{Figures \ref{nuofl}} --
\textbf{\ref{Gofl}} for three reasonable values of $b/a = 3,\: 2,\: \frac{3}{2}$.
The electron trajectory segment inside the cavity shown in \textbf{Figure
\ref{accunits}} was equal to the typical length of the pill-box cavity of SLAC
(4 cm). Calculations were performed for various values of $l$ and $m$. The optimal
parameters found were:
$a = 0.708$ cm, $b = 2.124$ cm, and $(l,\ m) = (6,\ -2)$. 

\section{Conclusions}

An electric field, intensified by structural resonance, can be used to accelerate
electrons. This is demonstrated here by placing a dielectric sphere
concentrically inside a spherical resonator, in which an appropriate whispering
gallery mode is excited. A strong, accelerating field appears next
to the surface of the dielectric. At the same time, the tangential component
of the magnetic field at the wall of the resonator is minimal. This makes losses
at the metallic walls negligible without engaging expensive cryogenic systems
ensuring superconductivity of the walls. The Q factor of the resonator only
depends on losses in the dielectric. For existing dielectrics, this gives a Q
factor three orders of magnitude better than obtained in existing cylindrical
cavities. Furthermore, for the proposed spherical cavity, \textit{all} field
components at the metallic wall are either zero or very small, see
Figure~\ref{fields}. Therefore, one can expect the proposed spherical cavity to be
less prone to electrical breakdowns than the traditional cylindrical cavity.

\begin{acknowledgments}
The authors would like to thank Professor Stanis{\l}aw Kuli\'nski for useful
discussions.
\end{acknowledgments}


\begin{thebibliography}{99}

\bibitem{wz}
W. {\.Z}akowicz, ``Whispering-Gallery-Mode Resonances: A New Way to Accelerate Charged Particles,'' \textit{Physical Review Letters},
Vol. 95, 2005, pp. 114801-114804.\\
doi:10.1103/PhysRevLett.95.114801

\bibitem{wza}
W. {\.Z}akowicz, ``Erratum: Whispering-Gallery-Mode Resonances: A New Way to
Accelerate Charged Particles [Phys. Rev. Lett. 95, 114801 (2005)],''
\textit{Physical Review Letters}, Vol. 97, 2006, p. 109901.\\
doi:10.1103/PhysRevLett.97.109901

\bibitem{wz1}
W. {\.Z}akowicz, ``Particle acceleration by wave scattering off dielectric spheres
at whispering-gallery-mode resonance,'' \textit{Physical
Review Special Topics--Accelerators and Beams}, Vol. 10, 2007, pp. 101301-101309.\\
doi:10.1103/PhysRevSTAB.10.101301

\bibitem{orni}
M. Ornigotti and A. Aiello, ``Theory of anisotropic whispering-gallery-mode resonators,'' \textit{Physical Review A}, Vol. 84, 2011, pp. 013828-013839.\\
doi:10.1103/PhysRevA.84.013828

\bibitem{lisey}
T. V. Liseykina, S. Pirner and D. Bauer, ``Relativistic Attosecond Electron Bunches
from Laser-Illuminated Droplets,'' \textit{Physical Review Letters},
Vol. 104, 2010, pp. 095002-095005.
doi:10.1103/PhysRevLett.104.095002

\bibitem{ilchen}
V. S. Ilchenko, A. A. Savchenkov, A. B. Matsko and L. Maleki, ``Nonlinear Optics and
Crystalline Whispering Gallery Mode Cavities,'' \textit{Physical Review Letters},
Vol. 92, 2004, pp. 043903-043906.\\
doi:10.1103/PhysRevLett.92.043903

\bibitem{taber}
R. C. Taber and C. A. Flory, ``Microwave oscillators incorporating
cryogenic sapphire dielectric resonators,'' \textit{IEEE Transactions on
Ultrasonics, Ferroelectrics and Frequency Control}, Vol. 42, 1995, pp. 111-119.\\
doi:10.1109/58.368306

\bibitem{krupka}
J. Krupka, K. Derzakowski, M. E Tobar,
J. Hartnett, and R. G. Geyer, ``Complex permittivity of some ultralow loss
dielectric crystals at cryogenic temperatures,'' \textit{Measurement Science and
Technology}, Vol. 10, 1999, pp. 387-392.\\
doi:10.1088/0957-0233/10/5/308

\bibitem{jackson}
J. D. Jackson, ``Classical electrodynamics,'' 3rd ed., John Wiley,
New York, 1998.

\bibitem{moroz}
A. Moroz, ``A recursive transfer-matrix solution for a dipole
radiating inside and outside a stratified sphere,'' \textit{Annals of Physics},
Vol. 315, 2005, pp. 352-418. doi:10.1016/j.aop.2004.07.002

\end{thebibliography}
\end{document}